\documentclass[conference]{IEEEtran}

\usepackage{amsmath,amsfonts}
\usepackage{graphicx}
\usepackage{cite}
\usepackage{booktabs}   

\usepackage[english]{babel}

\usepackage[letterpaper,top=2cm,bottom=2cm,left=3cm,right=3cm,marginparwidth=1.75cm]{geometry}

\usepackage{amsmath}
\usepackage{graphicx}
\usepackage[colorlinks=true, allcolors=blue]{hyperref}

\title{Privacy-Preserving Distributed Learning in IoT Systems: A Unified Threat Model and Evaluation Framework}
\author{
\IEEEauthorblockN{John Cartmell and Alexander Williams}
\IEEEauthorblockA{
Department of Electrical Engineering and Computer Science\\
Florida Atlantic University\\
Boca Raton, Florida, USA\\
jcartmell2023@fau.edu, alexanderwil2018@fau.edu
}
}

\begin{document}
\maketitle

\begin{abstract}
The increasing deployment of Internet-of-Things (IoT) devices has driven the adoption of distributed learning frameworks, where data remains local and model updates are shared across decentralized systems. While this paradigm reduces the need for centralized data collection, it introduces new privacy risks through the exchange of gradients, model parameters, and intermediate representations. A wide range of privacy-preserving techniques have been proposed to address these risks, including differential privacy, cryptographic methods, and system-level approaches. However, existing surveys typically describe these methods in isolation and lack a unified framework for evaluating their effectiveness under realistic attack models and resource constraints.

This paper presents a structured survey of privacy-preserving techniques for distributed learning with a focus on IoT environments. A unified threat model is introduced that captures key attack vectors, including model inversion, membership inference, gradient leakage, and communication-based attacks. Building on this model, an evaluation framework is developed to compare methods in terms of both privacy robustness and system-level efficiency, including computational, memory, and communication overhead.

Using this framework, representative techniques are analyzed, including distributed selective stochastic gradient descent, differential privacy, homomorphic encryption, secure multi-party computation, and Bloom Filter-based methods. The analysis highlights a fundamental trade-off between privacy strength and system efficiency, and reveals a broader design space spanning formal privacy mechanisms, cryptographic approaches, and lightweight probabilistic methods. In particular, Bloom Filter-based encodings are shown to occupy a distinct operating point in this space, introducing privacy through collision-induced ambiguity while maintaining low computational and communication overhead.

The results provide a unified view of privacy-preserving design choices for distributed learning in IoT systems and support informed decision-making based on application requirements and deployment constraints.
\end{abstract}

\section{Introduction}
The rapid growth of Internet-of-Things (IoT) devices has led to an increasing reliance on distributed learning paradigms, where data is generated and processed across a large number of decentralized and resource-constrained devices \cite{nguyen2021a, chataut2023}. In these systems, data is typically kept local to preserve privacy, while model updates or intermediate representations are shared to enable collaborative learning \cite{afzal2023, liu2022}. Although this approach reduces the need for centralized data collection, it introduces new privacy risks through the exchange of model parameters, gradients, and encoded data \cite{yaacoub2023}.

A range of privacy-preserving techniques have been proposed to address these challenges, including differential privacy \cite{geyer2018, pathak2010, wang2019}, cryptographic approaches such as homomorphic encryption and secure multi-party computation \cite{phong2018, bonawitz2017}, and various heuristic or system-level methods. While these techniques differ significantly in their underlying mechanisms, they are often evaluated in isolation, making it difficult to understand their relative strengths, limitations, and suitability for deployment in IoT environments \cite{saha2024, vyas2024}. In particular, existing approaches span fundamentally different design philosophies, ranging from formal privacy guarantees to computational confidentiality and lightweight data transformations that introduce ambiguity.

Existing survey papers typically provide descriptive overviews of these methods but often lack a unified framework for evaluating privacy guarantees and system-level trade-offs \cite{afzal2023, saha2024, nguyen2021}. In particular, there is limited emphasis on how different techniques perform under specific attack models, as well as how their computational, memory, and communication requirements impact their practicality in resource-constrained settings. As a result, the broader design space of privacy-preserving mechanisms, and the trade-offs between privacy strength and efficiency, remain insufficiently characterized.

To address these gaps, this paper presents a structured survey of privacy-preserving techniques for distributed learning with a focus on IoT environments. The key contribution of this work is the introduction of a unified threat model and evaluation framework that enables consistent comparison across methods. The threat model captures a range of adversarial capabilities, including model inversion, membership inference, gradient leakage, and communication-based attacks. The evaluation framework extends this analysis by incorporating system-level constraints such as computational overhead, memory usage, and communication cost.

Using this framework, we analyze a set of representative privacy-preserving techniques, including distributed selective stochastic gradient descent, differential privacy, homomorphic encryption, secure multi-party computation, and Bloom Filter-based methods \cite{cartmell2026bfml}. In addition, auxiliary approaches such as anonymization, blockchain-based coordination, and intrusion detection are considered in the context of their ability to enhance system-level privacy.

The results highlight a fundamental trade-off between privacy strength and system efficiency. Techniques that provide strong formal or cryptographic guarantees often incur significant overhead, limiting their applicability in IoT environments. In contrast, lightweight approaches introduce ambiguity or partial protection with substantially lower resource requirements. Bloom Filter-based methods, in particular, exemplify this class by encoding data into fixed-length representations that obscure feature-level information through controlled collisions while maintaining low computational and communication cost \cite{bloom1970, cartmell2026bfml}.

This work aims to provide a clear and structured understanding of this design space, enabling practitioners and researchers to make informed decisions when selecting privacy-preserving techniques for distributed learning systems. By grounding the analysis in a consistent threat model and evaluation framework, the paper moves beyond descriptive surveys and toward a more comparative and application-driven perspective.

\section{Background: IoT and Distributed Learning}
The integration of Internet-of-Things (IoT) devices with distributed learning has enabled large-scale data-driven applications across domains such as healthcare, smart infrastructure, and industrial systems \cite{nguyen2021a, chataut2023, jose2022}. In these environments, data is generated at the edge by numerous devices with limited computational, memory, and communication capabilities \cite{musaddiq2018, gupta2023}. Distributed learning frameworks allow these devices to collaboratively train models without requiring centralized data collection, reducing privacy risks associated with raw data aggregation \cite{afzal2023, liu2022}.

\begin{figure}
\centering
\includegraphics[width=\columnwidth]{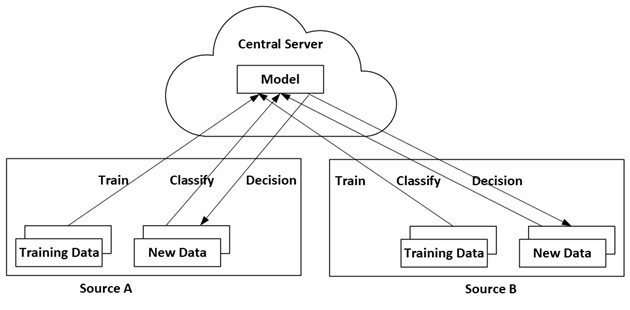}
\caption{Centralized distributed learning architecture where data from multiple sources is transmitted to a central server for training and inference.}
\label{fig:distributed_1_survey}
\end{figure}

A conventional centralized learning architecture is illustrated in Fig.~\ref{fig:distributed_1_survey}. In this paradigm, data generated at distributed sources is transmitted to a central server for training, classification, and decision-making. While this approach simplifies model management, it introduces significant communication overhead and raises privacy concerns due to the transmission of raw or minimally processed data \cite{afzal2023}.

\subsection{IoT System Constraints}
IoT devices are inherently resource-constrained. They typically operate with limited processing power, restricted memory capacity, and constrained network bandwidth \cite{musaddiq2018, gupta2023}. In addition, many IoT deployments rely on intermittent or unreliable connectivity, further limiting the feasibility of communication-intensive methods \cite{trindade}.

These constraints have a direct impact on the design of privacy-preserving techniques. Approaches that require heavy computation, large memory footprints, or frequent communication may not be practical in such environments \cite{afzal2023, yaacoub2023}. As a result, privacy mechanisms for IoT-based systems must balance protection with efficiency, ensuring that security enhancements do not exceed the capabilities of the underlying hardware. This constraint-driven setting naturally gives rise to a spectrum of approaches, ranging from computationally intensive methods with strong guarantees to lightweight transformations that trade formal guarantees for efficiency.

\subsection{Distributed Learning Paradigm}
Distributed learning enables multiple devices to collaboratively train a shared model while keeping data localized. In a typical setup, each device performs local training on its private dataset and periodically shares model updates, such as gradients or weights, with a central server or coordinating entity \cite{liu2022, afzal2023}. The server aggregates these updates to produce a global model, which is then redistributed to the devices.

\begin{figure}
\centering
\includegraphics[width=\columnwidth]{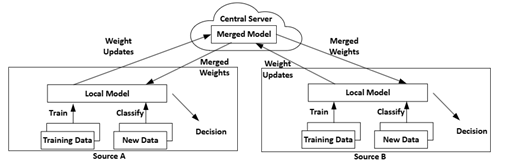}
\caption{Distributed learning architecture where local models are trained at each source and aggregated into a global model.}
\label{fig:distributed_2_survey}
\end{figure}

An alternative distributed learning paradigm is shown in Fig.~\ref{fig:distributed_2_survey}, where each device trains a local model using its private data and shares only model updates or weights with a central server. These updates are aggregated to form a global model, reducing the need to transmit raw data while enabling collaborative learning \cite{vyas2024, nguyen2021}.

While this paradigm reduces the need to transmit raw data, it introduces new privacy risks. Model updates can contain implicit information about the underlying data, which may be exploited by adversaries to reconstruct inputs or infer sensitive attributes \cite{shokri2015, yaacoub2023}. These risks are amplified in heterogeneous IoT environments, where devices may vary in trustworthiness and capability \cite{gupta2023}.

\subsection{Privacy Challenges in IoT-Based Distributed Learning}
The combination of IoT constraints and distributed learning introduces several key privacy challenges. First, the transmission of gradients or model updates creates opportunities for information leakage, even when raw data remains local \cite{yaacoub2023}. Second, adversaries may exploit repeated interactions over multiple training rounds to accumulate information about the underlying datasets \cite{saha2024}. Third, resource limitations restrict the use of computationally intensive privacy-preserving techniques, narrowing the set of feasible solutions and motivating the exploration of alternative, resource-efficient mechanisms \cite{nguyen2021}.

These challenges motivate the need for a structured evaluation of privacy-preserving methods that considers both adversarial capabilities and system constraints. In the following section, a unified threat model is introduced to characterize potential attacks and provide a consistent basis for comparing different techniques.

\section{Threat Model and Assumptions}
This section defines the system model, adversarial capabilities, and attack vectors considered in this survey. These definitions provide a consistent framework for evaluating privacy-preserving techniques in distributed learning systems deployed on resource-constrained Internet-of-Things (IoT) devices, and enable comparison across methods with fundamentally different design philosophies.

\subsection{System Model}
We consider a distributed learning architecture consisting of a set of $N$ IoT devices (clients) and a central aggregation server \cite{liu2022, afzal2023}. Each device collects local data and participates in collaborative model training without sharing raw data directly. Instead, devices either transmit model updates, transformed representations of data, or encrypted information to the central server.

The central server aggregates information received from participating devices to update a global model, which is subsequently redistributed to the devices for further training or inference \cite{vyas2024}. Communication between devices and the server may occur over unreliable or bandwidth-limited channels, reflecting typical IoT deployment constraints \cite{trindade}.

IoT devices are assumed to be resource-constrained, characterized by limited computational power, memory capacity, and communication bandwidth \cite{musaddiq2018, gupta2023}. These constraints play a critical role in determining the feasibility of privacy-preserving techniques.

\subsection{Adversary Model}
We consider multiple adversarial entities with varying capabilities and access levels.

\subsubsection{Honest-but-Curious Server}
The central server is assumed to follow the prescribed training protocol but may attempt to infer sensitive information from the data it receives. This includes analyzing model updates, gradients, or intermediate representations to reconstruct or approximate the original training data \cite{liu2022}.

\subsubsection{Malicious Client}
A participating client may deviate from the protocol by sending manipulated updates or attempting to infer information about other clients. Such a client may attempt to extract private data from the global model or from updates received during the training process \cite{saha2024, yaacoub2023}.

\subsubsection{External Eavesdropper}
An external adversary may observe communication between devices and the central server. This adversary does not participate in training but may attempt to reconstruct sensitive information from intercepted transmissions \cite{afzal2023}.

\subsection{Attack Model}
The following classes of privacy attacks are considered in evaluating the robustness of distributed learning systems.

\subsubsection{Model Inversion Attacks}
Model inversion attacks aim to reconstruct input data or sensitive features by exploiting access to model parameters or outputs \cite{shokri2015}. These attacks are particularly relevant when gradients or model weights are shared during training.

\subsubsection{Membership Inference Attacks}
Membership inference attacks attempt to determine whether a specific data sample was part of the training dataset \cite{shokri2015}. Such attacks can reveal sensitive participation information even when raw data is not directly exposed.

\subsubsection{Gradient Leakage and Reconstruction Attacks}
Gradient-based attacks exploit shared gradients or model updates to reconstruct original input data. Prior work has demonstrated that detailed information about training samples can be recovered from gradients under certain conditions \cite{geyer2018}.

\subsubsection{Communication Leakage}
Information may be leaked through transmitted data, including model updates, encrypted values, or transformed representations. Even when raw data is not shared, patterns in communication may expose sensitive information \cite{yaacoub2023, afzal2023}.

\subsection{Assumptions}
The following assumptions are made to bound the scope of analysis:
\begin{itemize}
    \item IoT devices operate under strict resource constraints, limiting the feasibility of computationally intensive cryptographic techniques \cite{musaddiq2018}.
    \item Communication channels may be insecure unless explicitly protected by a given method \cite{afzal2023}.
    \item The central server may be untrusted or only partially trusted \cite{liu2022}.
    \item Adversaries may have access to model updates, intermediate representations, or communication traffic, depending on the scenario.
\end{itemize}

\section{Evaluation Framework}
\label{sec:evaluation_framework}
To enable a consistent and meaningful comparison of privacy-preserving techniques, this section defines a structured evaluation framework based on the threat model introduced in the previous section. Rather than describing each method in isolation, the goal is to evaluate how well each approach mitigates specific privacy risks while remaining feasible for deployment on resource-constrained IoT devices.

The framework considers both \textit{privacy robustness} and \textit{system-level constraints}, allowing for a balanced assessment of each method and enabling comparison across approaches that differ in both underlying mechanisms and strength of guarantees \cite{afzal2023, nguyen2021}.

\subsection{Privacy Evaluation Criteria}
Privacy is evaluated based on resistance to the attack classes defined in the threat model. Each method is assessed with respect to the following criteria:

\subsubsection{Model Inversion Resistance}
This criterion measures the extent to which a method prevents reconstruction of input data from model parameters, outputs, or intermediate representations \cite{shokri2015}. Methods that obscure, transform, or otherwise limit the information content of shared representations achieve higher scores.

\subsubsection{Membership Inference Resistance}
This criterion evaluates whether an adversary can determine if a specific data sample was used during training \cite{shokri2015}. Techniques that introduce uncertainty or limit information leakage about individual samples provide stronger protection.

\subsubsection{Gradient Leakage Resistance}
This measures the ability of a method to prevent reconstruction of raw data from gradients or model updates. Prior work has demonstrated that gradients can reveal detailed information about training data under certain conditions \cite{geyer2018}. Approaches that avoid sharing gradients or sufficiently obfuscate them offer improved resistance.

\subsubsection{Communication Leakage Resistance}
This criterion considers whether sensitive information can be inferred from transmitted data, including model updates, encrypted values, or transformed representations \cite{yaacoub2023}. Methods that minimize, compress, or obfuscate transmitted information score higher.

Together, these criteria provide a structured view of how each method performs against realistic attack scenarios.

\subsection{Efficiency Metrics}
In addition to privacy, each method must be evaluated in terms of resource requirements, particularly in the context of IoT environments.

\subsubsection{Computational Overhead}
This metric captures the processing cost imposed on IoT devices. Methods requiring complex cryptographic operations or large-scale matrix computations incur higher computational overhead \cite{phong2018, bonawitz2017}.

\subsubsection{Memory Requirements}
This measures the amount of memory required to store model parameters, intermediate data, or auxiliary structures. Techniques that require large buffers or encrypted representations may be unsuitable for constrained devices \cite{musaddiq2018}.

\subsubsection{Communication Overhead}
This reflects the bandwidth required to transmit data between devices and the central server. Methods that involve large model updates or multiple communication rounds increase network load and may be impractical in low-bandwidth environments \cite{trindade, nguyen2021}.

\subsection{IoT Feasibility}
IoT feasibility captures whether a method can be realistically deployed given the constraints of typical IoT devices. While some techniques provide strong privacy guarantees, they may be impractical due to excessive computational or communication requirements \cite{afzal2023, yaacoub2023}.

A method is considered more suitable for IoT deployment if it:
\begin{itemize}
    \item operates within limited computational resources,
    \item requires minimal memory overhead,
    \item reduces communication bandwidth,
    \item and maintains acceptable model performance.
\end{itemize}

This criterion emphasizes practical deployability rather than theoretical capability.

\subsection{Scoring Methodology}
To enable comparison across methods, each technique is evaluated using a structured scoring approach based on the criteria defined above. Rather than relying on purely qualitative descriptions, scores are assigned to reflect relative performance across dimensions.

Each method is evaluated along two primary axes:
\begin{itemize}
    \item \textbf{Privacy Robustness:} Aggregated performance across the defined attack resistance criteria.
    \item \textbf{System Efficiency:} Combined assessment of computational, memory, and communication costs.
\end{itemize}

Scores are normalized to provide a consistent basis for comparison across methods with different scales and evaluation criteria. While the scoring remains semi-quantitative, it is grounded in reported characteristics from the literature, including computational complexity, communication requirements, and demonstrated resistance to known attacks \cite{afzal2023, saha2024}.

This approach allows methods with fundamentally different design philosophies---including formal privacy mechanisms, cryptographic protections, and lightweight transformations that introduce probabilistic ambiguity---to be compared within a unified framework. In particular, it highlights trade-offs between strong formal privacy guarantees and practical feasibility in constrained environments.

The resulting evaluation enables a clearer understanding of how different techniques perform under realistic deployment conditions, especially in IoT systems where resource limitations play a central role.

\section{Differential Privacy}
Differential Privacy (DP) is one of the most widely studied approaches for protecting sensitive information in distributed learning systems \cite{geyer2018, wang2019, saha2024}. The core idea is to introduce controlled randomness into data or model updates such that the contribution of any individual data point cannot be reliably distinguished. As a formal privacy mechanism, DP provides quantifiable guarantees under well-defined assumptions.

\subsection{Overview}
In distributed learning, differential privacy is typically applied by adding noise either to the local data, to gradients during training, or to aggregated model updates \cite{geyer2018, pathak2010}. This noise reduces the ability of an adversary to infer information about individual samples while preserving overall model utility.

\begin{figure}
\centering
\includegraphics[width=\columnwidth]{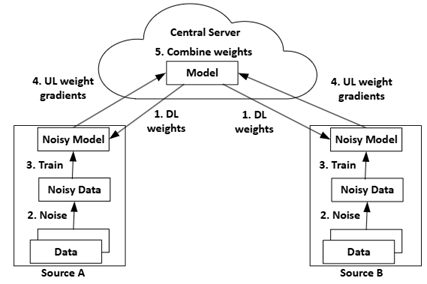}
\caption{Differential privacy in distributed learning. Noise is added to local data or model updates prior to training, producing a perturbed model that limits information leakage.}
\label{fig:dp_survey}
\end{figure}

The application of differential privacy in distributed learning is illustrated in Fig.~\ref{fig:dp_survey}. Each device perturbs its local data or model updates by adding noise prior to training or transmission. This process reduces the influence of individual data points and limits the ability of an adversary to reconstruct sensitive information from shared updates.

Several variants of DP have been proposed in the literature, including local differential privacy, where noise is added at the device level, and central differential privacy, where noise is added at the aggregation stage \cite{wang2019}. These approaches differ in terms of privacy guarantees and system overhead.

\subsection{Mathematical Formulation}
Differential privacy provides a formal privacy guarantee based on the concept of indistinguishability. A randomized mechanism $M$ satisfies $\epsilon$-differential privacy if for any two neighboring datasets $D$ and $D'$ that differ by a single data point, and for any output set $S$:

\begin{equation}
\Pr[M(D) \in S] \leq e^{\epsilon} \Pr[M(D') \in S]
\end{equation}

The parameter $\epsilon$ controls the strength of the privacy guarantee. Smaller values of $\epsilon$ provide stronger privacy but require the addition of more noise, which can degrade model performance \cite{saha2024}.

\subsection{Evaluation Under Threat Model}
Differential privacy provides strong protection against several attack classes defined in the threat model.

\subsubsection{Model Inversion Resistance}
By injecting noise into gradients or model updates, DP reduces the ability of an adversary to reconstruct input data from model parameters. However, the level of protection depends on the noise magnitude, and weak privacy settings may still allow partial reconstruction \cite{geyer2018}.

\subsubsection{Membership Inference Resistance}
DP is specifically designed to limit the influence of any single data point, making it difficult for an adversary to determine whether a particular sample was included in the training dataset \cite{shokri2015}. This is one of the strongest aspects of DP.

\subsubsection{Gradient Leakage Resistance}
When applied to gradients, DP reduces the risk of reconstructing raw data from shared updates. However, insufficient noise or improper implementation may still leave systems vulnerable to advanced gradient-based attacks.

\subsubsection{Communication Leakage Resistance}
DP does not inherently reduce the amount of information transmitted but instead perturbs the transmitted values. While this limits direct leakage, communication patterns and metadata may still expose information \cite{yaacoub2023}.

\subsection{Efficiency and IoT Feasibility}
While DP offers formal privacy guarantees, it introduces trade-offs that are particularly relevant in IoT environments.

\subsubsection{Computational Overhead}
The computational cost of DP is relatively low compared to cryptographic approaches. Noise generation and gradient clipping are generally lightweight operations, making DP feasible for many IoT devices \cite{afzal2023}.

\subsubsection{Memory Requirements}
DP does not significantly increase memory usage, as it operates directly on existing data or model updates. This makes it suitable for devices with limited storage capacity \cite{musaddiq2018}.

\subsubsection{Communication Overhead}
DP does not reduce communication size and may require additional rounds of training to compensate for noise-induced performance degradation. This can indirectly increase communication costs \cite{nguyen2021}.

\subsubsection{IoT Feasibility}
Overall, DP is well-suited for IoT environments due to its relatively low computational and memory requirements. However, achieving strong privacy guarantees may require careful tuning of privacy parameters, which can negatively impact model accuracy, especially in resource-constrained settings \cite{saha2024}.

\subsection{Summary}
Differential privacy provides a formal and well-understood privacy guarantee, making it a strong candidate for protecting sensitive data in distributed learning systems. It is particularly effective against membership inference attacks and offers meaningful protection against data reconstruction.

However, the trade-off between privacy and model utility remains a key limitation. In IoT environments, where resources are constrained and communication is limited, DP represents a practical middle ground between strong but computationally intensive cryptographic approaches and lightweight methods that rely on data transformation or ambiguity. Careful parameter selection is required to balance privacy, performance, and system overhead.

\section{Distributed Selective Stochastic Gradient Descent}
Distributed Selective Stochastic Gradient Descent (DSSGD) is an early approach to privacy-aware distributed learning that attempts to reduce information leakage by limiting the amount of data shared during training \cite{afzal2023, liu2022}. Unlike formal privacy-preserving methods, DSSGD does not provide explicit privacy guarantees but instead relies on partial information sharing to obscure sensitive data. As such, it represents a class of lightweight approaches that introduce privacy through reduced exposure rather than formal protection mechanisms.

\subsection{Overview}
In DSSGD, each participating device selectively downloads a subset of model parameters from a central server, performs local training using its private data, and then uploads only a subset of the computed gradients or weight updates. By restricting both the downloaded and uploaded information, DSSGD reduces the exposure of model parameters and intermediate computations.

The selection process is typically randomized, introducing additional uncertainty into which portions of the model are updated during each training round. This partial sharing mechanism aims to make it more difficult for an adversary to reconstruct the underlying training data, while also reducing communication overhead.

\begin{figure}
\centering
\includegraphics[width=\columnwidth]{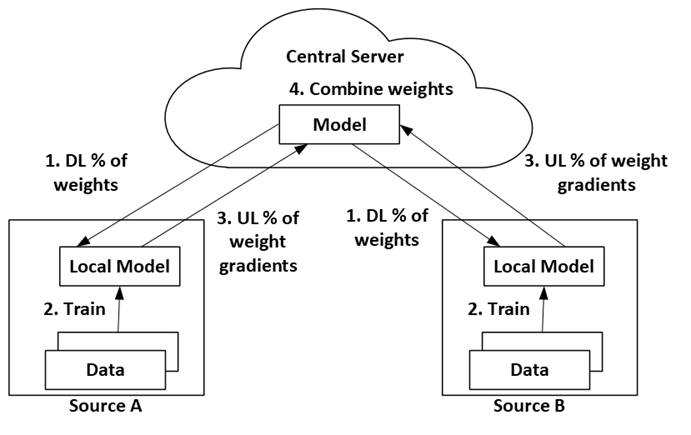}
\caption{Distributed Selective Stochastic Gradient Descent (DSSGD). Clients download a subset of model parameters, perform local training, and upload only partial gradients, reducing communication and limiting information exposure.}
\label{fig:dssgd_survey}
\end{figure}

The DSSGD workflow is illustrated in Fig.~\ref{fig:dssgd_survey}. Each client downloads only a subset of model parameters from the central server, performs local training using its private data, and uploads a subset of gradients or weight updates. The central server then aggregates these partial updates to update the global model. This selective communication reduces both bandwidth usage and the amount of information exposed during training.

\subsection{Evaluation Under Threat Model}
While DSSGD introduces some level of obfuscation, its effectiveness against the defined attack classes is limited.

\subsubsection{Model Inversion Resistance}
By sharing only a subset of gradients, DSSGD reduces the amount of information available to an adversary. However, the shared updates may still contain sufficient information to enable partial reconstruction of input data, particularly over multiple training rounds \cite{shokri2015}.

\subsubsection{Membership Inference Resistance}
DSSGD does not explicitly limit the influence of individual data points. As a result, an adversary may still infer whether specific samples contributed to the model, especially when observing repeated updates over time \cite{shokri2015}.

\subsubsection{Gradient Leakage Resistance}
Since DSSGD still transmits gradients, even if selectively, it remains vulnerable to gradient-based reconstruction attacks. The reduction in shared information may slow or complicate such attacks but does not eliminate the risk.

\subsubsection{Communication Leakage Resistance}
Selective sharing reduces the volume of transmitted data, which can limit direct exposure. However, patterns in the updates and repeated communication may still reveal information about the training data \cite{yaacoub2023}.

\subsection{Efficiency and IoT Feasibility}
One of the primary advantages of DSSGD is its relatively low system overhead.

\subsubsection{Computational Overhead}
DSSGD does not require complex cryptographic operations or additional transformations, resulting in low computational cost. This makes it suitable for devices with limited processing capabilities \cite{afzal2023}.

\subsubsection{Memory Requirements}
The method operates on subsets of model parameters and gradients, which can reduce memory usage compared to full-model updates. This is beneficial in constrained environments \cite{musaddiq2018}.

\subsubsection{Communication Overhead}
By transmitting only a fraction of gradients or parameters, DSSGD reduces communication bandwidth requirements. This is particularly advantageous in IoT systems with limited or intermittent connectivity \cite{trindade}.

\subsubsection{IoT Feasibility}
Due to its low computational and communication overhead, DSSGD is well-suited for deployment in IoT environments. However, this efficiency comes at the cost of weaker privacy protection compared to more advanced methods \cite{nguyen2021}.

\subsection{Summary}
DSSGD represents an early attempt to incorporate privacy considerations into distributed learning by limiting the amount of shared information. While it provides some degree of obfuscation through partial and randomized information exposure, it does not offer formal privacy guarantees and remains vulnerable to several attack vectors.

Its primary strength lies in its efficiency and suitability for resource-constrained devices. Within the broader design space, DSSGD can be viewed as a baseline lightweight approach that trades strong privacy guarantees for reduced communication and computational overhead. As more advanced techniques such as differential privacy and cryptographic methods have been developed, DSSGD serves as a useful reference point for understanding the trade-offs between partial information sharing, efficiency, and privacy protection.

\section{Homomorphic Encryption}
Homomorphic Encryption (HE) enables computation to be performed directly on encrypted data without requiring decryption \cite{phong2018, afzal2023}. In distributed learning systems, this property allows model updates to be aggregated while preserving the confidentiality of the underlying data.

\subsection{Overview}
In a typical distributed learning setting, each client computes local model updates and transmits them to a central server for aggregation. In HE-based approaches, these updates are encrypted before transmission. The server performs aggregation operations directly on the encrypted values and returns the result, which can then be decrypted by the clients \cite{phong2018}. Homomorphic encryption shifts the trust boundary by ensuring that sensitive data remains encrypted throughout the training and aggregation process.

\begin{figure}
\centering
\includegraphics[width=\columnwidth]{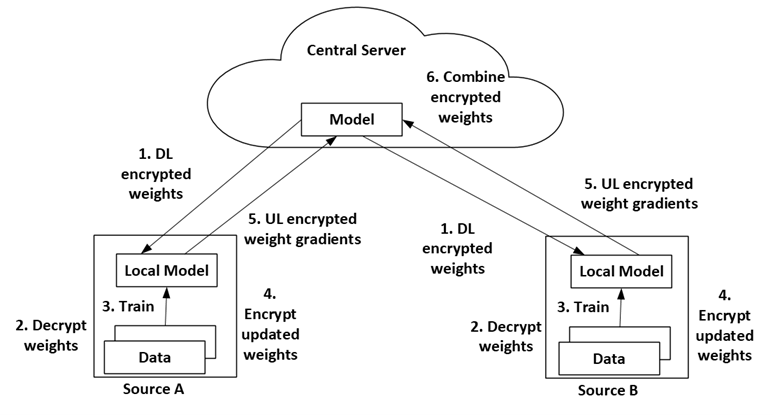}
\caption{Homomorphic encryption in distributed learning. Clients encrypt model updates before transmission, enabling secure aggregation at the server without exposing plaintext data.}
\label{fig:he_survey}
\end{figure}

The homomorphic encryption workflow in distributed learning is illustrated in Fig.~\ref{fig:he_survey}. Each client encrypts its local model updates prior to transmission, ensuring that the central server operates only on encrypted data. The server aggregates encrypted updates without decryption, and the resulting global model can be decrypted by the clients. This process preserves confidentiality of intermediate values while enabling collaborative training.

This process ensures that the server does not have access to the raw model updates in plaintext form, reducing the risk of information leakage from an honest-but-curious server \cite{afzal2023}. Variants of HE, including partially and fully homomorphic encryption schemes, differ in the types and number of operations that can be performed on encrypted data.

\subsection{Mathematical Formulation}
Homomorphic encryption schemes support operations on ciphertexts that correspond to operations on the underlying plaintexts. For an encryption function $\text{Enc}(\cdot)$ and decryption function $\text{Dec}(\cdot)$, an additive homomorphic property can be expressed as:

\begin{equation}
\text{Dec}(\text{Enc}(x_1) \oplus \text{Enc}(x_2)) = x_1 + x_2
\end{equation}

This property allows a central server to aggregate encrypted model updates without accessing the original values. More advanced schemes support both addition and multiplication, enabling more complex computations.

\subsection{Evaluation Under Threat Model}
Homomorphic encryption provides strong protection against several attack classes, particularly those involving access to intermediate data.

\subsubsection{Model Inversion Resistance}
Since model updates are encrypted during transmission and aggregation, the central server cannot directly analyze them to reconstruct input data. This significantly reduces the effectiveness of model inversion attacks from the server.

\subsubsection{Membership Inference Resistance}
HE limits access to individual updates, which reduces the ability of an adversary to determine whether a specific data point was included in training. However, if decrypted models or outputs are exposed, some risk may remain \cite{saha2024}.

\subsubsection{Gradient Leakage Resistance}
By encrypting gradients before transmission, HE prevents direct reconstruction of input data from intercepted updates. This provides strong protection against gradient-based attacks during communication.

\subsubsection{Communication Leakage Resistance}
HE protects the content of transmitted data but does not inherently hide communication patterns such as frequency or size of updates. As a result, side-channel information may still be exposed \cite{yaacoub2023}.

\subsection{Efficiency and IoT Feasibility}

Despite its strong privacy properties, HE introduces significant overhead that can limit its applicability in IoT environments.

\subsubsection{Computational Overhead}
Encryption and decryption operations in HE are computationally intensive, particularly for fully homomorphic schemes \cite{phong2018}. This can exceed the capabilities of many IoT devices with limited processing power \cite{nguyen2021}.

\subsubsection{Memory Requirements}
Encrypted representations are typically larger than their plaintext counterparts, increasing memory usage and storage requirements on both client devices and servers \cite{musaddiq2018}.

\subsubsection{Communication Overhead}
Ciphertexts generated by HE schemes are significantly larger than plaintext updates, resulting in increased bandwidth consumption \cite{nguyen2021}. This can be problematic in low-bandwidth or intermittently connected IoT systems.

\subsubsection{IoT Feasibility}
While HE provides strong confidentiality guarantees, its high computational and communication costs make it challenging to deploy directly on constrained IoT devices \cite{afzal2023}. In practice, HE is more suitable for systems where devices have sufficient resources or where computation can be offloaded.

\subsection{Summary}
Homomorphic encryption offers strong protection against data leakage by enabling secure computation over encrypted data. It is particularly effective against adversaries with access to intermediate model updates or communication channels.

However, the high computational, memory, and communication overhead associated with HE limits its practicality in IoT environments. As a result, HE is best suited for scenarios where strong privacy guarantees are required and sufficient resources are available, rather than for highly constrained systems. This positions homomorphic encryption at the high-security, high-overhead end of the privacy-efficiency spectrum considered in this work.

\section{Secure Multi-party Computation}
Secure Multi-party Computation (SMPC) enables multiple parties to collaboratively compute a function over their inputs while keeping those inputs private \cite{bonawitz2017, afzal2023}. In distributed learning systems, SMPC allows model training and aggregation to occur without exposing raw data or intermediate values to any single participant.

\subsection{Overview}
In SMPC-based distributed learning, each participating device splits its local data or model updates into multiple shares, which are distributed across different parties or computation nodes. These shares are constructed such that no single party can reconstruct the original data, but collectively they enable the computation of a desired function, such as gradient aggregation \cite{bonawitz2017}.

Unlike homomorphic encryption, which operates on encrypted values, SMPC relies on coordinated protocols between multiple parties to perform secure computation. These protocols ensure that intermediate values remain hidden throughout the computation process \cite{afzal2023}.

\subsection{Mathematical Formulation}
A common approach in SMPC is secret sharing. For example, a value $x$ can be split into $k$ shares $x_1, x_2, \dots, x_k$ such that:

\begin{equation}
x = \sum_{i=1}^{k} x_i
\end{equation}

Each share is distributed to a different party. Individually, the shares reveal no information about $x$, but collectively they allow reconstruction or computation of functions involving $x$.

Secure protocols enable operations such as addition and multiplication to be performed directly on shares without revealing the underlying values \cite{bonawitz2017}.

\subsection{Evaluation Under Threat Model}
SMPC provides strong privacy guarantees across multiple attack classes, particularly when assumptions about non-collusion are satisfied.

\subsubsection{Model Inversion Resistance}
Since no single party has access to complete model updates or raw data, SMPC significantly reduces the risk of reconstructing input data. Model inversion attacks are difficult to execute unless multiple parties collude.

\subsubsection{Membership Inference Resistance}
SMPC limits exposure of individual data contributions by distributing information across multiple parties. This reduces the likelihood that an adversary can determine whether a specific data point was used during training \cite{saha2024}.

\subsubsection{Gradient Leakage Resistance}
Because gradients are never revealed in full to any single entity, SMPC provides strong protection against gradient-based reconstruction attacks. Partial shares alone are insufficient to recover meaningful information.

\subsubsection{Communication Leakage Resistance}
While SMPC protects the content of shared data, it often requires multiple rounds of communication between parties. This increased interaction may expose communication patterns, although the underlying data remains protected \cite{yaacoub2023}.

\subsection{Efficiency and IoT Feasibility}
Despite its strong privacy guarantees, SMPC introduces significant overhead that can limit its practicality in IoT environments.

\subsubsection{Computational Overhead}
SMPC protocols require multiple rounds of coordinated computation, including secure operations on shared values. These operations are more complex than standard training procedures and can be computationally intensive for constrained devices \cite{afzal2023, nguyen2021}.

\subsubsection{Memory Requirements}
Maintaining multiple shares of data and intermediate values increases memory usage. This overhead can be challenging for IoT devices with limited storage capacity \cite{musaddiq2018}.

\subsubsection{Communication Overhead}
SMPC typically requires frequent communication between participating parties, often across multiple rounds per training step. This results in high communication overhead, which can be prohibitive in low-bandwidth or intermittently connected environments \cite{nguyen2021}.

\subsubsection{IoT Feasibility}
While SMPC provides strong privacy protection, its computational and communication requirements make it difficult to deploy directly on resource-constrained IoT devices \cite{afzal2023}. In practice, SMPC is more suitable for systems with sufficient infrastructure or where secure computation can be offloaded.

\subsection{Summary}
Secure Multi-party Computation offers strong privacy guarantees by ensuring that no single entity has access to complete data or model updates. It is highly effective against a wide range of attack vectors, including model inversion and gradient leakage.

However, the complexity and overhead of SMPC protocols limit their applicability in IoT environments. As a result, SMPC is best suited for scenarios where privacy requirements are strict and system resources are sufficient, rather than for highly constrained distributed systems. This places SMPC alongside homomorphic encryption at the high-privacy, high-overhead end of the design space considered in this work.

\section{Bloom Filter-Based Privacy Mechanisms}
Bloom Filters provide a probabilistic approach to data representation that can be leveraged to introduce privacy-preserving properties in distributed learning systems \cite{bloom1970, cartmell2026bfml}. Unlike formal privacy mechanisms such as differential privacy or cryptographic approaches, Bloom Filter-based methods rely on hashing and controlled collisions to obscure the original data.

\subsection{Overview}
A Bloom Filter is a fixed-length binary vector of size $m$ that represents a set of elements using $k$ independent hash functions \cite{bloom1970}. Each element is mapped to multiple positions in the vector, and the corresponding bits are set to 1. This process results in a compact representation that supports efficient membership queries with a controllable false positive rate.

In the context of distributed learning, Bloom Filters can be used as a preprocessing step to encode input features or data samples before training \cite{cartmell2026bfml}. Instead of transmitting raw data or gradients derived from raw data, devices operate on and share these encoded representations. This transformation introduces ambiguity due to hash collisions, making it difficult to reconstruct the original input. This distinguishes Bloom Filter-based approaches from both formal privacy mechanisms and cryptographic methods, as privacy is introduced through information compression and many-to-one mappings rather than noise injection or encryption.

\begin{figure}
\centering
\includegraphics[width=\columnwidth]{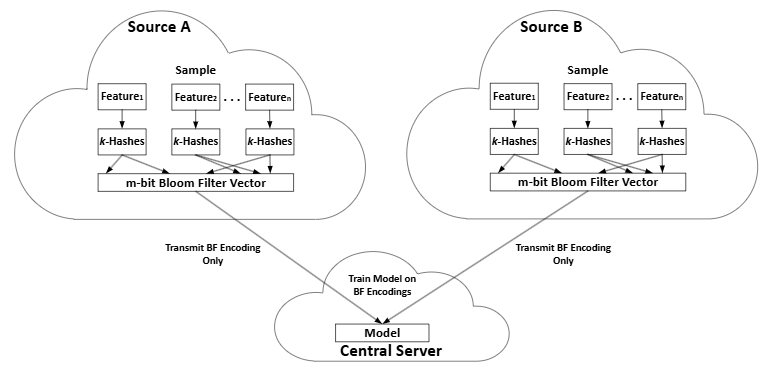}
\caption{Bloom Filter-based distributed learning pipeline. Each source encodes local data into a fixed-length $m$-bit Bloom Filter vector using $k$ hash functions. Only the encoded representations are transmitted to the central server, where model training is performed. This reduces communication overhead and limits direct exposure of raw feature values through collision-induced ambiguity.}
\label{fig:bf_survey}
\end{figure}

The Bloom Filter-based distributed learning pipeline is illustrated in Fig.~\ref{fig:bf_survey}. Each source encodes its local sample into a fixed-length Bloom Filter representation before transmission to the central server. Specifically, input features are mapped through $k$ hash functions into an $m$-bit Bloom Filter vector, which is then used for centralized model training. In this setting, only the encoded Bloom Filter representation is transmitted, while the raw feature values remain local to the source device. This reduces communication overhead and limits direct exposure of the original data, while introducing ambiguity through hash collisions.

Unlike differential privacy and cryptographic approaches, Bloom Filter encodings reduce data exposure by transforming feature representations prior to transmission rather than perturbing or encrypting them.

\subsection{Mathematical Formulation}
The false positive probability of a Bloom Filter after inserting $n$ elements into a filter of size $m$ using $k$ hash functions is given by:

\begin{equation}
P_{FP} = \left(1 - e^{-kn/m}\right)^k
\end{equation}

This probability reflects the likelihood that a query incorrectly indicates membership. In a privacy context, the same collision behavior that leads to false positives also contributes to ambiguity in the encoded representation, making inversion of the original data more difficult \cite{bloom1970}.

\subsection{Evaluation Under Threat Model}
Bloom Filter-based methods provide a different form of privacy compared to formal or cryptographic approaches.

\subsubsection{Model Inversion Resistance}
The use of hashing and collisions introduces a many-to-one mapping between input data and the Bloom Filter representation. This reduces the ability of an adversary to reconstruct original inputs, particularly when multiple features map to overlapping bit positions \cite{cartmell2026bfml}. However, this protection is probabilistic and depends on parameter selection.

\subsubsection{Membership Inference Resistance}
Bloom Filter encodings obscure individual feature values, making it more difficult to directly associate specific samples with model updates. While not designed explicitly to prevent membership inference, the loss of direct feature representation provides some level of protection.

\subsubsection{Gradient Leakage Resistance}
Since training is performed on encoded representations rather than raw data, gradient information corresponds to Bloom Filter features rather than original inputs. This reduces the direct interpretability of gradients and complicates reconstruction attacks, although it does not eliminate the possibility entirely.

\subsubsection{Communication Leakage Resistance}
Bloom Filters provide compact representations, reducing the amount of information transmitted compared to raw data or large model updates. Additionally, the encoded format obscures feature-level information, limiting direct leakage through communication \cite{yaacoub2023}.

\subsection{Efficiency and IoT Feasibility}
Bloom Filter-based approaches are particularly well-suited for IoT environments due to their efficiency.

\subsubsection{Computational Overhead}
The primary operations involved are hashing and bit manipulation, which are computationally lightweight. This makes Bloom Filters feasible for devices with limited processing capabilities \cite{musaddiq2018}.

\subsubsection{Memory Requirements}
Bloom Filters use fixed-size representations, allowing memory usage to be controlled explicitly through the choice of $m$. This provides predictable and often low memory overhead.

\subsubsection{Communication Overhead}
Encoded representations are compact and fixed in size, reducing bandwidth requirements compared to transmitting raw data or high-dimensional model updates \cite{trindade}.

\subsubsection{IoT Feasibility}
Due to their low computational and communication requirements, Bloom Filter-based methods are highly suitable for deployment in resource-constrained IoT environments \cite{afzal2023}. They provide a practical alternative when more computationally intensive privacy techniques are not feasible.

\subsection{Summary}
Bloom Filter-based methods offer a lightweight and practical approach to introducing privacy in distributed learning systems. By leveraging hashing and collision-induced ambiguity, they provide resistance to direct data reconstruction while maintaining low system overhead.

Unlike differential privacy or cryptographic methods, Bloom Filters do not provide formal privacy guarantees. However, they occupy a useful position in the design space by offering a balance between privacy, efficiency, and deployability, particularly in IoT environments where resource constraints are a primary concern \cite{cartmell2026bfml}.

\section{Auxiliary and System-Level Privacy Enhancements}
In addition to the primary privacy-preserving techniques discussed in previous sections, several approaches aim to enhance privacy at the system or preprocessing level \cite{afzal2023, nguyen2021}. These methods do not provide standalone privacy guarantees under the defined threat model but can complement core techniques in practical deployments.

Unlike differential privacy or cryptographic methods, these approaches typically address specific aspects of the system, such as data preprocessing, trust management, or anomaly detection. As a result, their effectiveness depends on how they are integrated with other privacy-preserving mechanisms.

\subsection{Anonymization Techniques}
Anonymization involves removing or transforming personally identifiable information (PII) from datasets prior to training. Common approaches include generalization, suppression, and perturbation of sensitive attributes.

While anonymization can reduce direct exposure of sensitive data, it is generally insufficient against modern attacks such as model inversion and membership inference. Adversaries can often exploit auxiliary information or statistical correlations to re-identify individuals, especially in high-dimensional datasets \cite{shokri2015}.

From a system perspective, anonymization introduces minimal computational and communication overhead, making it suitable for IoT environments \cite{musaddiq2018}. However, due to its limited robustness under the defined threat model, anonymization is best used as a preprocessing step in combination with stronger privacy-preserving techniques.

\subsection{Blockchain-Based Approaches}
Blockchain-based methods focus on improving trust, transparency, and accountability in distributed learning systems \cite{trindade}. These approaches use decentralized ledgers to record model updates, verify participant behavior, and manage access control.

While blockchain can help mitigate issues such as data tampering and unauthorized participation, it does not inherently provide privacy protection for the underlying data or model updates. Sensitive information may still be exposed unless combined with techniques such as encryption or differential privacy \cite{afzal2023}.

Additionally, blockchain systems often introduce significant computational and communication overhead, which can be challenging for resource-constrained IoT devices \cite{nguyen2021}. As a result, blockchain is more appropriately viewed as a coordination and trust mechanism rather than a direct privacy solution.

\subsection{Intrusion Detection Systems}
Intrusion Detection Systems (IDS) aim to identify malicious or anomalous behavior within distributed learning environments. These systems may use machine learning models to detect clients that attempt to poison the model or deviate from expected behavior \cite{yaacoub2023}.

IDS can enhance system security by identifying adversarial participants, thereby reducing the risk of indirect privacy breaches caused by malicious actors. However, IDS does not prevent information leakage from legitimate model updates or communication channels.

In terms of system requirements, IDS typically introduces moderate computational and communication overhead. While feasible in some IoT deployments, its effectiveness depends on the availability of sufficient data and monitoring infrastructure \cite{afzal2023}.

\subsection{Additive and Multiplicative Masking Schemes}
Additive and multiplicative masking techniques aim to obscure data or model updates by combining them with random values before transmission. In some approaches, multiple parties share components of these random values to enable secure aggregation.

These methods can provide a degree of protection against direct inspection of transmitted data. However, they often rely on assumptions about non-collusion or the secrecy of masking values. If these assumptions are violated, the underlying data may be exposed \cite{bonawitz2017}.

From an efficiency standpoint, masking schemes are generally less computationally intensive than full cryptographic methods but may still introduce communication overhead depending on the protocol design \cite{nguyen2021}. Their privacy guarantees are typically weaker than those provided by formal or cryptographic approaches.

\subsection{Summary}
Auxiliary and system-level techniques play an important role in enhancing privacy and security in distributed learning systems. However, they do not provide comprehensive protection against the attack classes defined in the threat model when used in isolation.

These methods are most effective when combined with core privacy-preserving techniques, contributing to a layered approach that balances privacy, efficiency, and system robustness \cite{afzal2023}.

\section{Comparative Analysis and Scoring}
This section applies the evaluation framework defined earlier to compare the primary privacy-preserving techniques. The goal is to provide a structured view of how each method performs under the defined threat model while considering system-level constraints relevant to IoT environments \cite{afzal2023, saha2024}. :contentReference[oaicite:0]{index=0}

\subsection{Method Comparison Across Threat Model}
Each method is evaluated based on its resistance to the attack classes defined in Section~\ref{sec:evaluation_framework}. Table~\ref{tab:threat_comparison} summarizes the relative performance of each method.

\begin{table*}[t]
\centering
\caption{Comparison of Methods Under Threat Model}
\label{tab:threat_comparison}
\begin{tabular}{lcccc}
\hline
\textbf{Method} & \textbf{Model Inversion} & \textbf{Membership Inference} & \textbf{Gradient Leakage} & \textbf{Communication Leakage} \\
\hline
DSSGD & Low & Low & Low & Medium \\
Differential Privacy & Medium--High & High & Medium & Medium \\
Homomorphic Encryption & High & High & High & High \\
SMPC & High & High & High & High \\
Bloom Filters & Medium & Medium & Medium & Medium--High \\
\hline
\end{tabular}
\end{table*}

DSSGD provides limited protection, as it relies primarily on reducing the amount of shared information rather than formally securing it \cite{afzal2023}. Differential privacy introduces noise to limit the influence of individual data points, offering strong protection against membership inference but only moderate protection against reconstruction attacks depending on parameter selection \cite{geyer2018, wang2019}.

Homomorphic encryption and SMPC provide strong protection across all attack classes by ensuring that intermediate data is never exposed in plaintext form \cite{phong2018, bonawitz2017}. However, these methods rely on cryptographic or multi-party protocols that introduce significant system overhead \cite{nguyen2021}.

Bloom Filter-based methods provide moderate protection by introducing ambiguity through hashing and collisions. While they do not offer formal guarantees, they reduce the direct interpretability of data and gradients, particularly in resource-constrained environments \cite{bloom1970, cartmell2026bfml}.

\subsection{Efficiency and Resource Trade-offs}
In addition to privacy robustness, each method is evaluated in terms of computational cost, memory usage, and communication overhead. Table~\ref{tab:efficiency_comparison} summarizes these characteristics.

\begin{table*}[t]
\centering
\caption{Efficiency and Resource Comparison}
\label{tab:efficiency_comparison}
\begin{tabular}{lccc}
\hline
\textbf{Method} & \textbf{Computation} & \textbf{Memory} & \textbf{Communication} \\
\hline
DSSGD & Low & Low & Low \\
Differential Privacy & Low--Medium & Low & Medium \\
Homomorphic Encryption & High & Medium--High & High \\
SMPC & High & High & High \\
Bloom Filters & Low & Low & Low \\
\hline
\end{tabular}
\end{table*}

DSSGD and Bloom Filter-based approaches exhibit low computational and communication overhead, making them suitable for constrained environments \cite{afzal2023, musaddiq2018}. Differential privacy introduces moderate overhead due to noise addition and potential increases in training iterations \cite{saha2024}.

In contrast, homomorphic encryption and SMPC incur substantial computational and communication costs due to encryption operations and multi-party coordination \cite{phong2018, bonawitz2017}. These requirements can exceed the capabilities of typical IoT devices \cite{nguyen2021}.

\subsection{IoT Feasibility Analysis}
Combining privacy robustness with system efficiency provides insight into the practical applicability of each method in IoT environments.

DSSGD and Bloom Filter-based methods are well-suited for deployment on resource-constrained devices due to their low overhead \cite{musaddiq2018}. However, DSSGD offers limited privacy protection, while Bloom Filters provide a balance between efficiency and moderate privacy \cite{cartmell2026bfml}.

Differential privacy represents a practical compromise, offering formal privacy guarantees with manageable system overhead \cite{geyer2018, saha2024}. Its effectiveness depends on careful tuning of privacy parameters to balance privacy and model performance.

Homomorphic encryption and SMPC provide strong privacy guarantees but are generally impractical for direct deployment on constrained IoT devices \cite{nguyen2021}. These methods are better suited for systems with greater computational resources or hybrid architectures where heavy computation can be offloaded.

\subsection{Overall Trade-off Analysis}
The comparison highlights a clear trade-off between privacy strength and system efficiency, consistent with prior observations in distributed learning and IoT systems \cite{afzal2023, nguyen2021}.

Methods such as homomorphic encryption and SMPC provide strong protection across all attack classes but require significant computational and communication resources. Differential privacy offers a middle ground, providing formal guarantees with moderate overhead.

Bloom Filter-based approaches occupy a distinct position in this design space as a lightweight, ambiguity-driven privacy mechanism. While they do not provide formal privacy guarantees, they introduce privacy through information compression and collision-induced many-to-one mappings, which reduce the direct interpretability of data and gradients \cite{bloom1970}. Combined with their low computational and communication overhead, this makes them particularly attractive for IoT systems where resource constraints are a primary concern \cite{cartmell2026bfml}.

\subsection{Summary}
The results demonstrate that no single method is optimal across all dimensions. Instead, the choice of technique depends on the specific requirements of the system, including the desired level of privacy, available computational resources, and communication constraints.

For IoT environments, where efficiency is critical, lightweight approaches such as Bloom Filters and selectively applied differential privacy provide practical solutions \cite{saha2024}. In contrast, cryptographic and multi-party methods are more appropriate for settings where strong privacy guarantees are required and sufficient resources are available \cite{nguyen2021}.

\section{Conclusion}
This paper presented a structured survey of privacy-preserving techniques for distributed learning in Internet-of-Things (IoT) environments. Unlike traditional surveys that describe methods in isolation, this work introduced a unified threat model and evaluation framework to enable consistent comparison across techniques.

The analysis considered multiple classes of adversaries and attack vectors, including model inversion, membership inference, gradient leakage, and communication-based attacks. Each method was evaluated not only in terms of privacy robustness but also with respect to computational, memory, and communication constraints that are critical in IoT systems \cite{afzal2023, nguyen2021}.

The results highlight a fundamental trade-off between privacy strength and system efficiency. Cryptographic approaches such as homomorphic encryption and secure multi-party computation provide strong protection across all attack classes but impose significant computational and communication overhead \cite{phong2018, bonawitz2017}. Differential privacy offers formal guarantees with more manageable resource requirements, though it introduces a trade-off between privacy and model performance \cite{geyer2018, saha2024}.

Bloom Filter-based methods occupy a distinct position within this design space. While they do not provide formal privacy guarantees, they offer a lightweight and efficient mechanism for introducing ambiguity into data representations \cite{bloom1970}. This makes them particularly suitable for resource-constrained environments where more computationally intensive techniques may be impractical \cite{cartmell2026bfml}.

Overall, the findings indicate that no single approach is optimal for all scenarios. Instead, the selection of a privacy-preserving technique must be guided by the specific requirements of the application, including the desired level of privacy, available system resources, and deployment constraints.

Future work includes the development of hybrid approaches that combine lightweight data transformations with formal privacy mechanisms, as well as the exploration of adaptive methods that dynamically balance privacy and efficiency based on system conditions. Further investigation into parameter selection and theoretical characterization of probabilistic privacy mechanisms also remains an open area of research.


\bibliographystyle{IEEEtran}
\bibliography{references}

\end{document}